\documentclass[lettersize,journal,twoside]{IEEEtran} 
\ifCLASSINFOpdf
  \usepackage[pdftex]{graphicx}
\else
\fi
\usepackage{tikz}
\usetikzlibrary{patterns}
\usepackage{pgfplots}
\pgfplotsset{compat=newest}
\usetikzlibrary{plotmarks}
\usetikzlibrary{arrows.meta}
\usepgfplotslibrary{patchplots}
\usepackage{grffile}
\usepackage{cite}
\usepackage{balance}

\usepackage{amsmath,amssymb,amsfonts}
\usepackage{mathtools} 
\usepackage{amsthm}
\usepackage{lipsum}
\usepackage{cuted}
\usepackage[bottom]{footmisc}

\usepackage{tabularx}
\usepackage{booktabs}
\usepackage[caption=false]{subfig} %

\makeatletter
\let\oldfootnote\footnote
\def\footnote{\@ifstar\footnote@star\footnote@nostar}
\def\footnote@star#1{{\let\thefootnote\relax\footnotetext{#1}}}
\def\footnote@nostar{\oldfootnote}
\makeatother
\usepackage{array}
\usepackage[nolist]{acronym}
\begin{acronym}
\acro{AoA}{angle of arrival}
\acro{AoD}{angle of departure}
\acro{AWGN}{additive white Gaussian noise}
\acro{BER}{bit error rate}
\acro{BS}{base station}
\acro{CFAR}{constant false alarm rate}
\acro{CP}{cyclic prefix}
\acro{CRLB}{Cramér-Rao lower bound}
\acro{EFIM}{equivalent Fisher information}
\acro{FAR}{false-alarm rate}
\acro{FMCW}{frequency modulated continuous wave}
\acro{FIM}{Fisher information matrix}
\acro{GLRT}{generalized likelihood ratio test}
\acro{ICI}{inter-carrier interference}
\acro{IoT}{Internet of Things}
\acro{ISAC}{integrated sensing and communication}
\acro{ISFFT}{inverse symplectic finite Fourier transform}
\acro{IFFT}{inverse fast Fourier transform}
\acro{FFT}{fast Fourier transform}
\acro{ISI}{inter-symbol interference}
\acro{JSC}{joint sensing and communication}
\acro{LoS}{line of sight}
\acro{LS}{least squares}
\acro{MIMO}{multiple-input multiple-output}
\acro{ML}{maximum likelihood}
\acro{OFDM}{orthogonal frequency-division multiplexing}
\acro{OTFS}{orthogonal time frequency space}
\acro{PER}{packet error rate}
\acro{PSD}{power spectral density}
\acro{QAM}{quadrature amplitude modulation}
\acro{QPSK}{quadrature phase shift keying}
\acro{RCS}{radar cross-section}
\acro{PEB}{position error bound}
\acro{RMSE}{root mean square error}
\acro{Rx}{receiver}
\acro{RoI}{region of interest}
\acro{SFFT}{symplectic finite Fourier transform}
\acro{SNR}{signal-to-noise ratio}
\acro{Tx}{transmitter}
\acro{UE}{user equipment}
\acro{ULA}{uniform linear array}
\acro{V2V}{vehicle-to-vehicle}
\acro{V2I}{vehicle-to-infrastructure}
\acro{V2X}{vehicular-to-everything}
\acro{5G}{fifth generation}
\end{acronym}

\usepackage{xcolor}
\newcommand{\Jm}{\mathbf{J}_\mathrm{m}}
\newcommand{\Jn}{\mathbf{J}_\mathrm{n}}
\newcommand{\Nbs}{N_\mathrm{bs}}

\newcommand{\transp}{\mathsf{T}}
\newcommand{\herm}{\mathsf{H}}
\DeclareMathOperator{\Tr}{Tr}
 
\newcommand{\Nr}{N_{\mathrm{R}}}
\newcommand{\Nt}{N_{\mathrm{T}}}
\newcommand{\av}{\mathbf{a}}

\newcommand{\bv}{\mathbf{b}}
\newcommand{\w}{\mathbf{w}_\mathrm{T}}
\newcommand{\hm}{\mathbf{h}}

\newcommand{\I}{\boldsymbol{\mathcal{I}}}

\newcommand{\ytilde}{\mathbf{\tilde{y}}}
\newcommand{\y}{\mathbf{y}}
\newcommand{\Thetab}{\boldsymbol{\Theta}}
\newcommand{\thetab}{\boldsymbol{\theta}}

\begin{document}
\title{Cooperative Maximum Likelihood Target Position Estimation for MIMO-ISAC Networks}
\author{\IEEEauthorblockN{
Lorenzo~Pucci,~\IEEEmembership{Member,~IEEE}, Tommaso~Bacchielli,~\IEEEmembership{Graduate~Student~Member,~IEEE}, and Andrea~Giorgetti,~\IEEEmembership{Senior~Member,~IEEE}
\thanks{© 2025 IEEE. Personal use of this material is permitted. Permission from IEEE must be obtained for all other uses, in any current or future media, including reprinting/republishing this material for advertising or promotional purposes, creating new collective works, for resale or redistribution to servers or lists, or reuse of any copyrighted component of this work in other works. DOI: 10.1109/LWC.2025.3548446.}
\thanks{This work was supported by the European Union - Next Generation EU under the Italian National Recovery and Resilience Plan (NRRP), Mission 4, Component 2, Investment 1.3, partnership on ``Telecommunications of the Future'' (PE00000001 - program ``RESTART''). The authors are with the Wireless Communications Laboratory, CNIT, DEI, University of Bologna, Italy, 
Email: \{lorenzo.pucci3, tommaso.bacchielli2, andrea.giorgetti\}@unibo.it
}}
}
\markboth{}%
{Pucci \MakeLowercase{\textit{et al.}}: Cooperative Maximum Likelihood Target Position Estimation for MIMO-ISAC Networks}

\maketitle

\begin{abstract}
This letter investigates target position estimation in integrated sensing and communication networks composed of multiple cooperating monostatic base stations (BSs). Each BS employs a MIMO-orthogonal time-frequency space (OTFS) scheme, enabling the coexistence of communication and sensing. A general cooperative maximum likelihood (ML) framework is derived, directly estimating the target position in a common reference system rather than relying on local range and angle estimates at each BS. Positioning accuracy is evaluated in single-target scenarios by varying the number of collaborating BSs, using root mean square error (RMSE), and comparing against the square root of the Cramér-Rao lower bound. Numerical results demonstrate that the ML framework significantly reduces the position RMSE as the number of cooperating BSs increases.
\end{abstract}


\IEEEpeerreviewmaketitle

\acresetall
\section{Introduction}
\IEEEPARstart{T}{he} marriage of \ac{ISAC} and mobile networks unlocks sensing capabilities not yet available, while also facilitating cooperative multi-\acp{BS} strategies that have proven effective for improved localization and tracking \cite{CoopISAC:M24}. Recent studies have examined \ac{BS} selection and resource allocation in cooperative \ac{ISAC} systems \cite{Xu_CoopSensing:C23, Wei_symbolLevel_coopISAC:J24, Gao_CoopISAC:J23, Chen2024jointnodeselectionresource, MengMasouros:C24}, yet tracking high-speed targets (e.g., autonomous vehicles) remains challenging under severe Doppler effects and rapidly changing environments. Among candidate waveforms, \ac{OTFS} modulation is robust to Doppler shifts and maps time-varying channels to the delay-Doppler domain, making it well-suited for \ac{ISAC} \cite{Gaudio2020}. Fully exploiting these advantages requires processing frameworks that leverage the spatial diversity inherent in cooperative multi-\acp{BS} systems. However, a key challenge persists in deriving a closed-form expression for cooperative \ac{ML} target position estimation from jointly processed signals at multiple \acp{BS}.

To address this, we propose an \ac{OTFS}-based cooperative \ac{ISAC} system with multiple monostatic \ac{MIMO} \acp{BS} and a general \ac{ML} framework for target position estimation. Specifically, in the proposed framework, signals from each \ac{BS} are processed at a fusion center to estimate the target position relative to a common reference system. Additionally, we derive the \ac{CRLB} for position estimation in the proposed cooperative system, whose square root, the \ac{PEB}, provides a theoretical lower bound on accuracy. Numerical results demonstrate that \ac{BS} cooperation based on \ac{ML} estimation significantly improves localization accuracy—often achieving multi-fold improvements compared to single \ac{BS} configurations—and validate the \ac{ML} estimation against the \ac{PEB}.

In this letter, capital and lowercase bold letters denote matrices and vectors. 
The operators $(\cdot)^\transp$, $(\cdot)^\herm$, $\Tr(\cdot)$, and $\|\cdot\|$ refer to the transpose, conjugate transpose, trace, and Euclidean norm; $\{a_i\}^{N}_{i=1}$ denotes a vector $\mathbf{a} = [a_1, \dots,a_N]$ and $\mathbf{I}_n$ is the $n \times n$ identity matrix; $\mathbb{E}\{\cdot\}$ and $\mathbb{V}\{\cdot\}$ denote the mean value and variance; $\mathbf{x} \sim \mathcal{CN}(\mathbf{0},\boldsymbol{\Sigma})$ represents a zero-mean circularly symmetric complex Gaussian vector with covariance $\boldsymbol{\Sigma}$. Additionally, $\mathfrak{Re}(\cdot)$, $|\cdot|$, and $\angle(\cdot)$ denote the real part, the absolute value, and the argument of a (real or complex) number, and $\otimes$ represents the Kronecker product.

The letter is organized as follows: Section~\ref{sec:system_model} introduces the system model, Section~\ref{sec:sensing_estimation} outlines the cooperative ML estimation framework and the \ac{CRLB}, Section~\ref{sec:num_res} presents numerical results, and Section~\ref{sec:conclusion} concludes the paper.

\section{System Model} \label{sec:system_model}

\subsection{Physical Model} \label{sec:physic-mod}
\begin{figure}[t]
    \centering
    \includegraphics[width=0.99\linewidth]{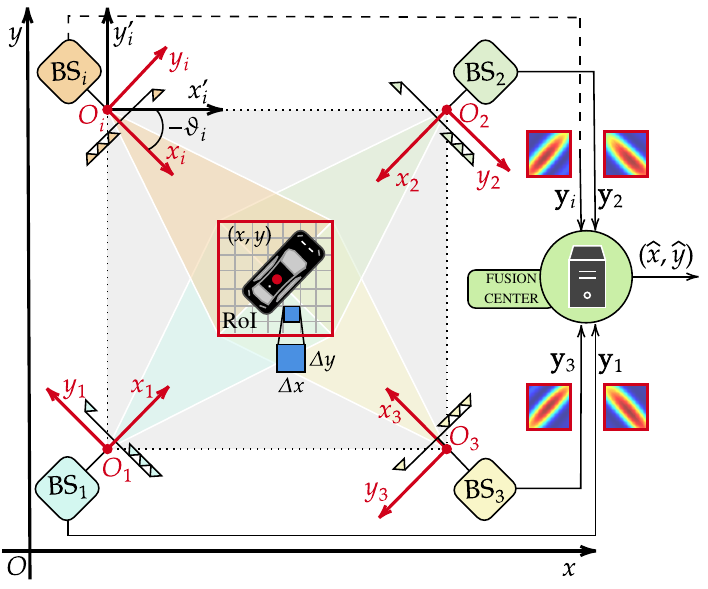}
     \caption{Network of $\Nbs$ cooperating \acp{BS} for target localization. The red square marks the \ac{RoI} considered for fine-grained grid-based \ac{ML} estimation.}
    \label{fig:ISAC_OTFS_system_model}
\end{figure}

We consider an \ac{OTFS}-based network, where monostatic \acp{BS}, each equipped with an \ac{ISAC} transceiver with $\Nt$ transmit and $\Nr$ receive antennas, cooperate through a fusion center to localize targets within a designated region, as shown in Fig.~\ref{fig:ISAC_OTFS_system_model}. Each \ac{BS} can split sensing and communication in a time- or frequency-division manner based on the required quality of service.\footnote{Interference between \acp{BS} can be mitigated using frequency and time division approaches, possibly complemented by dynamic resource allocation methods such as dynamic frequency reuse and inter-cell resource sharing \cite{yaugciouglu2022dynamic}.}

The network operates at a carrier frequency $f_\mathrm{c}$, with a bandwidth $B = M \Delta f \ll f_\mathrm{c} $, where $M$ is the number of subcarriers and $\Delta f = 1/T$ is the subcarrier spacing, with $T$ being the time slot duration.
Consider a scenario with $P$ point-like targets, each moving with a radial velocity $v_p$ at a distance $r_p$ from the monostatic transceiver and with \ac{RCS} $\sigma_p$. Assuming far-field \ac{LoS} propagation between the transceiver and each target, the $\Nr \times \Nt$ complex baseband channel impulse response of the radar function is 
\begin{equation} \label{eqn:h_channel}
    \mathbf{H}(t,\tau) = \sum_{p=0}^{P-1} \alpha_p \bv (\phi_p) \av^\herm (\phi_p) \delta(\tau-\tau_p) e^{j2\pi f_{\mathrm{D},p} t}
\end{equation}
where $\alpha_p$ is the complex channel gain, accounting for attenuation and phase shift, $f_{\mathrm{D},p}=2v_p f_\mathrm{c}/c$ represents the Doppler shift, and $\tau_p=2r_p/c$ is the round-trip delay, with $c$ denoting the speed of light. According to the radar equation, $|\alpha_{p}|^2~= \frac{G^2\sigma_p \, c^2}{(4 \pi)^3 \, f_\mathrm{c}^2 \, r_{p}^4}$, where $G^2$ accounts for the single antenna element gain $G$ of the \ac{Tx} and \ac{Rx} arrays.
Furthermore, $\av(\phi_p)$ and $\bv(\phi_p)$ denote the \ac{Tx} and \ac{Rx} array response vectors associated with the $p$-th target, with $\phi_p$ the $p$-th \ac{AoD}/\ac{AoA}, assumed to be equal in the monostatic setup. For \acp{ULA} with $\Nt$ antenna elements spaced at half-wavelength intervals, the array response vector is given by $\mathbf{a}(\phi_p) = [e^{-j\frac{\Nt-1}{2}\pi \sin \phi_p}, \dots, e^{j\frac{\Nt-1}{2}\pi \sin \phi_p}]^\transp$. The same definition applies to $\mathbf{b}(\phi_p)$ with $\Nr$.

\subsection{OTFS-ISAC Input-Output Relationship} \label{sec:OTFS-in-out-rel}
Considering a total of $N$ time slots, the vector of signals transmitted through the $\Nt$ antennas at the $n$-th time slot is
\begin{equation} \label{eqn:s}
    \mathbf{s}_n(t) = \sqrt{P_\mathrm{avg}}\sum_{m=0}^{M-1} X[n,m] \w e^{j2\pi m\Delta f (t -nT)} g_{\mathrm{tx}}(t-nT)
\end{equation}
where $P_\mathrm{avg} = P_\mathrm{T}/M$ is the average transmit power per subcarrier with $P_\mathrm{T}$ the total transmit power dedicated to sensing. The vector $\w$ represents the unit-norm beamforming vector, chosen to provide nearly constant gain over a given circular sector, following the approach outlined in \cite{FriedlanderMIMOradar}. The pulse shape is represented by $g_\mathrm{tx}(t)$ and $X[n,m] \in \mathbb{C}$ is the complex symbol located in the $N \times M$ time-frequency domain grid obtained via  \ac{SFFT} from complex data symbols $x[k,l] \in \mathcal{A}$, where $\mathcal{A}$ is the complex alphabet and $\mathbb{E}\{|x[k,l]|^2\}=1$, arranged in the $M \times N$ delay-Doppler domain grid, i.e.,
\begin{equation} \label{eqn:X_JSC}
    X[n,m] = \frac{1}{\sqrt{NM}}\sum_{k=0}^{M-1} \sum_{l=0}^{N-1} x[k,l] e^{-j2\pi(\frac{mk}{M}-\frac{nl}{N})}
\end{equation}
\noindent with $n,l=0,...,N-1$ and $m,k=0,...,M-1$. Further details on the processing chain are available in \cite{Hadani2018, Gaudio2020, BacPucPaoGio:J24}.

The transmitted signal $\{\mathbf{s}_n(t)\}^{N-1}_{n=0}$, after the channel \eqref{eqn:h_channel}, is collected at each \ac{Rx} antenna and processed by a filter matched to the demodulating pulse $g_{\mathrm{rx}}(t)$. In the ideal case of bi-orthogonality between transmitting and receiving pulses for integer multiples of $T$ and $\Delta f$, \ac{ISI} can be avoided without a \ac{CP}. However, in the more realistic scenario of imperfect bi-orthogonality, as considered here, some residual \ac{ISI} occurs without a \ac{CP} \cite{Hadani2018}. 
The output of the matched filter is then sampled at $t = n T$ and $f = m \Delta f$, forming a $N \times M$ symbol grid in the time-frequency domain, whose $(n,m)$ element is \cite{Hadani2018}
\begin{equation} \label{eqn:Y_symbol}
    Y[n,m] = \sum_{n'=0}^{N-1} \sum_{m'=0}^{M-1} X[n',m'] \hm_{n,m}[n',m']
\end{equation}
where $ \hm_{n,m}[n',m']$ is the $\Nr \times 1$ channel vector in the time-frequency domain. Without loss of generality, we focus on one target, i.e., $P=1$ in \eqref{eqn:h_channel}, so the index $p$ is dropped and the channel vector becomes
\begin{align}
    \hm_{n,m}[n',m'&] = \\ & h \bv (\phi) e^{j2\pi(n'Tf_{\mathrm{D}}-m\Delta f\tau)} A_{g_{\mathrm{r}}, g_{\mathrm{t}}}\left( \tau_{n,n'},f_{\mathrm{D},m,m'}\right)\nonumber
    \label{eq:channel_vec}
\end{align}
where $h \triangleq \sqrt{P_\mathrm{avg}}\, \gamma  \, \alpha \, e^{j2\pi f_{\mathrm{D}} \tau}$ is the overall complex channel factor, being $\gamma  = \av^\herm (\phi) \w$ the complex beamforming factor related to the \ac{AoD} of the target, $\tau_{n,n'} = (n - n')T - \tau$, and $f_{\mathrm{D},m,m'} = (m - m')\Delta f - f_{\mathrm{D}}$.
The term $A_{g_{\mathrm{r}}, g_{\mathrm{t}}}\left( \tau_{n,n'},f_{\mathrm{D},m,m'}\right)$ is the cross-ambiguity function, in the delay-Doppler domain, between the two pulses $g_{\mathrm {r}}(t)$ and $g_{\mathrm {t}}(t)$, here both chosen to be rectangular of duration $T$.\footnote{The cross-ambiguity function between two generic pulses $g_u(t)$ and $g_v(t)$ of duration $T$ is defined as \cite{Gaudio2020}
\begin{align*}
A_{g_u,g_v}(\tau,f_{\mathrm{D}}) & \triangleq \int_{0}^{T} g_u(t) g_v^\ast(t-\tau) e^{-j2\pi f_{\mathrm{D}} t} dt.
\end{align*}}
Lastly, the demodulated symbols, back in the delay-Doppler domain, can be obtained by performing \ac{ISFFT} on the symbols in \eqref{eqn:Y_symbol}, as 
$\mathbf{y}[k,l] = h \bv (\phi) \sum_{k'=0}^{M-1} \sum_{l'=0}^{N-1} x[k',l'] \Psi_{l,l'}[k,k']$,
where $\Psi_{l,l'}[k,k']$ is the $(l,l',k,k')$ complex element of the matrix $\mathbf{\Psi} \in \mathbb{C}^{MN \times MN}$, which retains information on Doppler shift and round-trip delay for the target, defined as
\begin{equation*} \label{eq:Psi}
    \Psi_{l,l'}[k,k'] = \frac{1}{NM} \sum_{n,n',m,m'} \Lambda_{n,n',m,m'} A_{g_{\mathrm{r}},g_{\mathrm{t}}}( \tau_{n,n'}, f_{\mathrm{D},m,m'})
\end{equation*}
where $\Lambda_{n,m,n',m'} =e^{j2\pi\left(n'Tf_{\mathrm{D}}-m\Delta f \tau-\frac{nl-n'l'}{N}+\frac{mk-m'k'}{M}\right)}$. Under the hypothesis of rectangular pulses and a maximum channel delay smaller than $T$, an easy-to-handle expression of this matrix's elements can be found in \cite{Gaudio2020, BacPucPaoGio:J24}.

Through a vectorization operation performed on the transmitted symbols $x[k,l]$, the input-output relationship 
can be written compactly as
\begin{equation} \label{eqn:y_hat_ISAC_vect}
    \mathbf{y} = h \mathbf{G}(f_{\mathrm{D}},\tau,\phi) \mathbf{x} + \mathbf{\boldsymbol{\nu}}
\end{equation}
where $\mathbf{y} \in \mathbb{C}^{M N \Nr \times 1}$ contains the received symbols at each Rx antenna, $\mathbf{G}(f_{\mathrm{D}},\tau,\phi) \triangleq \bv (\phi) \otimes \boldsymbol{\Psi}\in \mathbb{C}^{M N \Nr \times M N}$ is the effective channel matrix\cite{Dehkordi_TWC23}, $\mathbf{\boldsymbol{\nu}} \sim \mathcal{CN}(\mathbf{0}_{M N \Nr}, \sigma_\nu^2 \mathbf{I}_{M N \Nr})$ is the complex \ac{AWGN} vector with $\sigma_\nu^2 = N_0 \Delta f$, with $N_0$ the noise \ac{PSD}.

\section{Sensing Parameters Estimation and Cramér-Rao Lower Bound} \label{sec:sensing_estimation}
\subsection{Cooperative Maximum Likelihood Position Estimation} \label{sec:cooperative_ML}
Let us consider the scenario in Fig.~\ref{fig:ISAC_OTFS_system_model}, where $\Nbs$ \acp{BS} at coordinates $\boldsymbol{\mathcal{O}}_i = [x^{(i)}_{\mathrm{bs}}, y^{(i)}_{\mathrm{bs}}]^\transp$, with $i=1,\dots,\Nbs$, cooperate to locate the point-like target at $\mathbf{p} = [x,y]^\transp$ in a common Cartesian reference system.  The target position in the local reference system centered at $\boldsymbol{\mathcal{O}}_i$ is given by $\mathbf{p}_i = [x_{i}, y_{i}]^\transp = [r_i \cos(\phi_i), r_i \sin(\phi_i)]^\transp$, with $r_i = \tau_i\,c/2$. 

By collecting the backscattered signals by the $i$-th \ac{BS}, $\y_i$, detailed in \eqref{eqn:y_hat_ISAC_vect}, the vector $\ytilde \in \mathbb{C}^{M N \Nr \Nbs \times 1}$ of all received symbols is 
\begin{equation}
\label{eq:ytilde}
\ytilde=\bigl[\y_1^\transp,\y_2^\transp, \dots, \y^\transp_{\Nbs}\bigr]^\transp.
\end{equation}
%
The vector of target parameters to be estimated with respect to the $i$-th \ac{BS} is $\thetab_i = [\beta_i, \varphi_i, f_{\mathrm{D},i}, \tau_i,  \phi_i]^\transp$, where $\beta_i = |h_i|$ and $\varphi_i~=~\angle h_i$, and the complete set of unknown parameters is indicated with 
%
$\Thetab = \bigl[\thetab_1^\transp, \thetab_2^\transp \cdots, \thetab^{\transp}_{\Nbs}\bigr]^\transp \in \Gamma$,
%
where $\Gamma = \mathbb{C}^{\Nbs} \times \mathbb{R}^{3 \Nbs}$ is the considered parameter space.

Ignoring the terms that are not relevant for the estimation, the log-likelihood function of the received signal in \eqref{eq:ytilde} is
%
%
%
\begin{equation}
    \label{eq:log-likelihood}
       l(\ytilde;\Thetab,\mathbf{x}) \approx -\frac{1}{\sigma^2_\nu}\sum_{i=1}^{\Nbs} \bigl\| \y_i - h_i \mathbf{G}(f_{\mathrm{D},i},\tau_i,\phi_i) \mathbf{x}_i  \bigr\|^2.
\end{equation}
From \eqref{eq:log-likelihood}, an \ac{ML} estimation problem can be formulated to estimate the unknown parameters in $\Thetab$, as follows
\begin{equation}
\label{eq:ML_estim_multi}
\widehat{\Thetab}_\mathrm{ML} = \underset{\Thetab \in \Gamma}{\arg \max} \; l(\tilde{\y};\Thetab,\mathbf{x}).
\end{equation}

Notably, the estimate of each complex channel coefficient $h_i$ depends solely on the received signal $\mathbf{y}_i$ at the $i$-th \ac{BS}. Under this premise, the \ac{ML} estimator of the $i$-th channel coefficient, $\widehat{h}_i$, is given by \eqref{eq:h_estim} in the Appendix. By substituting $\widehat{h}_i$ into \eqref{eq:log-likelihood} and applying the same steps as outlined in the Appendix, \eqref{eq:ML_estim_multi} can be rewritten as
\begin{equation}
\label{eq:ML_estim_multi_2}
\{\widehat{f}_{\mathrm{D},i}, \widehat{\tau}_i, \widehat{\phi}_i\}^{\Nbs}_{i=1} = \hspace{-0.3cm} \underset{\{f_{\mathrm{D},i}, \tau_i, \phi_i\}^{\Nbs}_{i=1} \in \Omega}{\arg\max} \; \sum_{i=1}^{\Nbs}\frac{|\y_i^\herm \mathbf{G}(f_{\mathrm{D},i},\tau_i,\phi_i) \mathbf{x}_i|^2}{\bigl\|\mathbf{G}(f_{\mathrm{D},i},\tau_i,\phi_i)\, \mathbf{x}_i\bigr\|^2}
\end{equation}
where $\Omega = \mathbb{R}^{3 \Nbs}$ is the search space. The \ac{ML} estimation in \eqref{eq:ML_estim_multi_2} can be further simplified as derived below.

Referring to Fig.~\ref{fig:ISAC_OTFS_system_model}, the relationship between $\mathbf{p}_i$ and $\mathbf{p}$ is given by the following transformation
\begin{align}
\label{eq:coord_transform}
\left\{x_i = x'_i\cos(\vartheta_i) + y'_i\sin(\vartheta_i);
y_i = -x'_i\sin(\vartheta_i) + y'_i\cos(\vartheta_i)\right\}
\end{align}
where $x'_i=x-x^{(i)}_\mathrm{bs}$ and $y'_i=y-y^{(i)}_\mathrm{bs}$, while $\vartheta_i$ is the counterclockwise rotation angle of the $i$-th reference system with respect to the common one. The local coordinates of the target, $\mathbf{p}_i$, are in turn related to the unknown parameters $(\tau_i, \phi_i)$, through 
\begin{equation}
\label{eq:cartesian_to_polar}
\Bigl\{\,\phi_i=\arctan(y_{i}/x_{i});\,\, \tau_i=\frac{2}{c}\sqrt{x_{i}^2+y_{i}^2}\,\Bigr\}.
\end{equation}
Therefore, the parameters $\{\tau_i, \phi_i\}^{\Nbs}_{i=1} \subset \boldsymbol{\Theta}$ can be expressed as a function of the target position $\mathbf{p}$ in the common reference system, i.e., $\{\tau_i(\mathbf{p})$, $\phi_i(\mathbf{p})\}^{\Nbs}_{i=1}$.
The \ac{ML} estimator in \eqref{eq:ML_estim_multi_2} can then be rewritten as 
\begin{align}
\label{eq:ML_pos_estim}
\bigl[ & \widehat{f}_{\mathrm{D},1}, \dots, \widehat{f}_{\mathrm{D},\Nbs}, \widehat{\mathbf{p}}\bigr] = \\ & = \underset{\left[\{f_{\mathrm{D},i}\}^{\Nbs}_{i=1}, \mathbf{p} \right] \in \Pi}{\arg\max} \; \sum_{i=1}^{\Nbs}\frac{|\y_i^\herm \mathbf{G}(f_{\mathrm{D},i},\tau_i(\mathbf{p}),\phi_i(\mathbf{p})) \mathbf{x}_i|^2}{\bigl\|\mathbf{G}(f_{\mathrm{D},i},\tau_i(\mathbf{p}),\phi_i(\mathbf{p}))\, \mathbf{x}_i\bigr\|^2}\nonumber
\end{align}
with $\Pi = \mathbb{R}^{\Nbs + 2}$ the new search space.
\begin{table} [t] 
\centering
 \caption{JSC network parameters} \label{tab:sim_param}
 \resizebox{0.99\columnwidth}{!}{
 \renewcommand{\arraystretch}{1}
 \begin{tabular}{l | c | l}
\toprule
Parameters & Symbols & Values \\
\midrule
Number of Tx and Rx antennas & $\Nt$, $\Nr$ & $16$ \\
Antenna element gain & $G$ & $1$\\
Tx beamwidth / Beamforming gain & $\mathrm{BW}$ / $|\gamma|^2$ & $40^\circ$ / $\sim 4.1$ dB \\
Total transmit power & $P_\mathrm{T}$ & $30$ dBm \\
Carrier frequency & $f_\mathrm{c}$ & $60$ GHz\\
Time slots / Active subcarriers & $N$ / $M$ & $50$ / $96$\\
Subcarrier spacing / Time slot duration & $\Delta f$ / $T$ & $1$ MHz / $1\,\mu$s \\
Noise PSD / Target RCS & $N_0$ / $\sigma$  & $4\cdot 10^{-20}$ W/Hz / $1$ m$^2$\\
Region-of-interest / pixel size & RoI / $(\Delta x \times \Delta y)$ & $16$ m$^2$ / $(0.02 \times 0.02)$ m$^2$ \\
 \bottomrule
\end{tabular}
}
\label{NRparam}
\end{table}

Lastly, if for a given $\mathbf{p}$, or $(\tau_i, \phi_i)$ pair, the estimate of the Doppler shift, $\widehat{f}_{\mathrm{D},i}$, can be computed at each \ac{BS} (e.g., through a coarse estimation procedure or by exploiting information from tracking algorithms) and then replaced in \eqref{eq:ML_pos_estim}, the cooperative \ac{ML} target position estimation for an \ac{ISAC} network composed of $\Nbs$ \ac{BS} can be computed as
\begin{equation}
\label{eq:ML_pos_estim_2}
\widehat{\mathbf{p}} = \underset{\mathbf{p} \in \mathbb{R}^2}{\arg\max} \; \sum_{i=1}^{\Nbs}\frac{|\y_i^\herm \mathbf{G}(\widehat{f}_{\mathrm{D},i},\tau_i(\mathbf{p}),\phi_i(\mathbf{p})) \mathbf{x}_i|^2}{\bigl\|\mathbf{G}(\widehat{f}_{\mathrm{D},i},\tau_i(\mathbf{p}),\phi_i(\mathbf{p}))\, \mathbf{x}_i\bigr\|^2}.
\end{equation}
Interestingly, \eqref{eq:ML_pos_estim_2} also applies to \ac{OFDM}, where the effective channel matrix $\mathbf{G}$ can be represented as in \cite[Equation (23)]{DehPucJunGioPaoCai:J24}.
\subsection{Numerical Evaluation of Cooperative Maximum Likelihood} \label{sec:numerical_ML}
As mentioned, solving \eqref{eq:ML_pos_estim_2} requires an initial estimation of the Doppler shifts $\{f_{\mathrm{D},i}\}^{\Nbs}_{i=1}$, achieved by evaluating \eqref{eq:ML_estim_multi_2} within the defined search space $\Omega$. For a cooperative \ac{ISAC} system with
$\Nbs \geq 2$ \acp{BS}, this \ac{ML} estimation operates in a continuous search space of dimension $\geq 6$, leading to considerable complexity. To simplify computation, we employ an approximate method: breaking down the estimation into $\Nbs$ independent problems, each in a $3$D search space $\tilde\Omega~\subset~\Omega$.
 
Assuming a prior target detection phase has identified a \ac{RoI} as in \cite{DehPucJunGioPaoCai:J24}, the search space for $\{\tau_i, \phi_i\}^{\Nbs}_{i=1}$ in \eqref{eq:ML_estim_multi_2} is limited to range of values within the \ac{RoI} (see Fig.~\ref{fig:ISAC_OTFS_system_model}).\footnote{The number of targets and their \acp{RoI} can be determined through target detection methods, such as \ac{GLRT} or \ac{CFAR} detectors (see \cite{Xu_CFAR_detector:J23} and references therein).} Doppler shifts $\{f_{\mathrm{D},i}\}^{\Nbs}_{i=1}$ are similarly restricted, based on the scenario.
Hence, we define a discrete set of tuples $(f_{\mathrm{D},i}, \tau_i, \phi_i) \in \tilde\Omega$ as
   $$\tilde\Omega=\{f_{\mathrm{D},i}^\mathrm{min},...,f_{\mathrm{D},i}^\mathrm{max}\}\!\times\!\{
   \tau_i^\mathrm{min},...,\tau_i^\mathrm{max}\}\!\times\!
   \{\phi_i^\mathrm{min},...,\phi_i^\mathrm{max}\}$$
with steps $\delta f_\mathrm{D} = c_{f_\mathrm{D}} \Delta f_\mathrm{D}$, $\delta \tau = c_\tau \Delta \tau$, and $\delta \phi = c_\phi \Delta \Phi$, being $\Delta f_\mathrm{D} =1/(NT)$ and $\Delta \tau = 1/(M\Delta f)$ the system resolution in Doppler and delay, respectively, and $\Delta \Phi$ the beamwidth at $0\,$degrees. Moreover, $c_{f_\mathrm{D}}, c_\tau, c_\phi \in (0,1]$ are factors used to control the desired accuracy of parameter estimation.
Once Doppler shifts $\{\widehat{f}_{\mathrm{D},i}\}^{\Nbs}_{i=1}$ are estimated, \eqref{eq:ML_pos_estim_2} can then be used to compute $\widehat{\mathbf{p}}$. As before, the search is discretized within the \ac{RoI} by defining a discrete set of common coordinates as $\{x_{\mathrm{min}},...,x_{\mathrm{max}}\}\!\times\! \{y_{\mathrm{min}},...,y_{\mathrm{max}}\}$, with steps $\Delta x$ and $\Delta y$, respectively, chosen to achieve the desired accuracy. Notably, \eqref{eq:ML_pos_estim_2} can be interpreted as a summation of $\Nbs$ radar maps in the $x$-$y$ plane, computed within the \ac{RoI}, as shown in Fig.~\ref{fig:ISAC_OTFS_system_model}.
\subsection{Cramér-Rao Lower Bound Derivation}\label{sec:CRLB}
This section derives the \ac{CRLB}, the fundamental lower limit on the variance of unbiased estimators, specifically for target position estimation.

Let $\boldsymbol{\mu}_i =~ h_i\, \mathbf{G}(f_{\mathrm{D},i},\tau_i,\phi_i)\, \mathbf{x}_i \in \mathbb{C}^{M N \Nr \times 1}$ be the vector of mean values of the received signal $\mathbf{{y}}_i$ at the $i$-th \ac{BS}. For the vector of unknown parameters $\boldsymbol{\theta}_i$, the \ac{CRLB} is given by
\begin{equation} \label{eqn:Fisher_var}\nonumber
    \mathbb{V}\{{\hat{\theta}_{i,q}}\} \geq  \mathrm{CRLB}(\theta_{i,q})= \left[\I^{-1}(\thetab_i)\right]_{{q,q}}\; q=1,\dots,\mathrm{card(\boldsymbol{\theta}_i)}
\end{equation}
where ${\hat{\theta}_{i,q}}$ is the estimate of the $q$-th parameter contained in $\boldsymbol{\theta}_i$, and $[\I^{-1}(\thetab_i)]_{{q,q}}$ represents the $q$-th element on the main diagonal of the inverse of the \ac{FIM} $\I(\thetab_i)$. The vector $\boldsymbol{\mu}_i$ can be reshaped into a $3$D array with elements $\mu_{i}[k,l,j]$, indexed by $(k, l, j)$, corresponding to dimensions $(M\times~N~\times~\Nr)$, respectively. The generic $qp$-th element of $\I(\thetab_i)$ can then be computed as \cite{BacPucPaoGio:J24}
\begin{equation} \label{eqn:Fisher}
    \left[\I(\thetab_i)\right]_{{q,p}} = \frac{2}{\sigma^2_{\nu}} \mathfrak{Re} \Biggl\{ \sum_{k,l,j} \biggl(\frac{\partial \mu_{i}[k,l,j]}{\partial \theta_{i,q}}\biggr)^\ast \biggl(\frac{\partial \mu_i[k,l,j]}{\partial \theta_{i,p}}\biggr) \Biggl\}
\end{equation}
where
\begin{equation} \label{eq:s_n}
    \mu_{i}[k,l,j] = \beta_i e^{j \varphi_i }b_j(\phi_i)\sum_{k'=0}^{M-1}\sum_{l'=0}^{N-1}  \Psi_{l,l'}[k,k'] x_i[k',l']
\end{equation}
with $b_j(\phi)$ the $j$-th element of the array response vector $\mathbf{b}(\phi)$.\footnote{As can be seen by substituting \eqref{eq:s_n} into \eqref{eqn:Fisher}, solving \eqref{eqn:Fisher} requires computing the partial derivatives of $\mathbf{\Psi}$, whose closed-form expressions can be found in \cite{BacPucPaoGio:J24}.} 
The vector ${\boldsymbol{\theta}}_i$ can be partitioned as ${\boldsymbol{\theta}}_i=[\boldsymbol{\theta}_{i,1}^\transp,\boldsymbol{\theta}_{i,2}^\transp]^\transp$, with $\boldsymbol{\theta}_{i,1}=[\beta_i, \varphi_i, f_{\mathrm{D},i}]^\transp$, $\boldsymbol{\theta}_{i,2}=[\tau_i,\phi_i]^\transp$, which induces the partition of the \ac{FIM} into the matrices $\mathbf{A}\in \mathbb{R}^{3\times 3}$, $\mathbf{B}\in \mathbb{R}^{3\times 2}$, and $\mathbf{C}\in \mathbb{R}^{2\times 2}$, as
%
$\I(\boldsymbol{\theta}_i)~=~
\big[\begin{smallmatrix}
\mathbf{A} & \mathbf{B} \\
\mathbf{B}^\transp & \mathbf{C}
\end{smallmatrix}\big].$ Then, the \ac{EFIM} is \cite{shen2010fundamental1, PucGio2024PEB}: 
$\I_\mathrm{e}(\boldsymbol{\theta}_{i,2})=\mathbf{C}-\mathbf{B}^\transp\mathbf{A}^{-1}\mathbf{B}.$

Considering that each \ac{BS} independently observes the target parameters, the \ac{FIM} for cooperative position estimation is given by \cite{shen2010fundamental1, PucGio2024PEB}:
  $\I_\mathrm{e}(\mathbf{p}) = \sum_{i=1}^{N_\mathrm{BS}} \Jn^\mathsf{T} \Bigl(\Jm^\transp \I_\mathrm{e}(\boldsymbol{\theta}_{i,2})\Jm \Bigr)\Jn$,
where $\Jm$ and $\Jn$ represent the Jacobian of the transformations $\mathbf{p}_i~\rightarrow~(\tau_i,\phi_i) $, and $\mathbf{p}~\rightarrow~ \mathbf{p}_i$, respectively, whose definition, considering \eqref{eq:cartesian_to_polar} and \eqref{eq:coord_transform}, are as follows
\begin{equation}
\label{eq:EFIM_pos}
\Jm \triangleq
\begin{bmatrix}
 \frac{\partial \tau_i}{\partial x_i} & \frac{\partial \tau_i}{\partial y_i} \\
 \frac{\partial \phi_i}{\partial x_i} & \frac{\partial \phi_i}{\partial y_i}
\end{bmatrix},
\qquad
\Jn \triangleq
\begin{bmatrix}
\frac{\partial x_i}{\partial x} & \frac{\partial x_i}{\partial y} \\
\frac{\partial y_i}{\partial x} & \frac{\partial y_i}{\partial y}
\end{bmatrix}.
\end{equation}

After computing \eqref{eq:EFIM_pos}, the \ac{CRLB} of position estimate is determined as $\mathbb{V}\{\hat{\mathbf{p}}\} \geq \mathrm{CRLB}(\mathbf{p})=\Tr(\I_\mathrm{e}^{-1}(\mathbf{p}))$. The \ac{PEB} is then defined as
$\mathrm{PEB} \triangleq \sqrt{\mathrm{CRLB}(\mathbf{p})}$.
\section{Numerical Results} \label{sec:num_res}
\begin{figure}[t]
     \centering
     \includegraphics[width=1\linewidth]{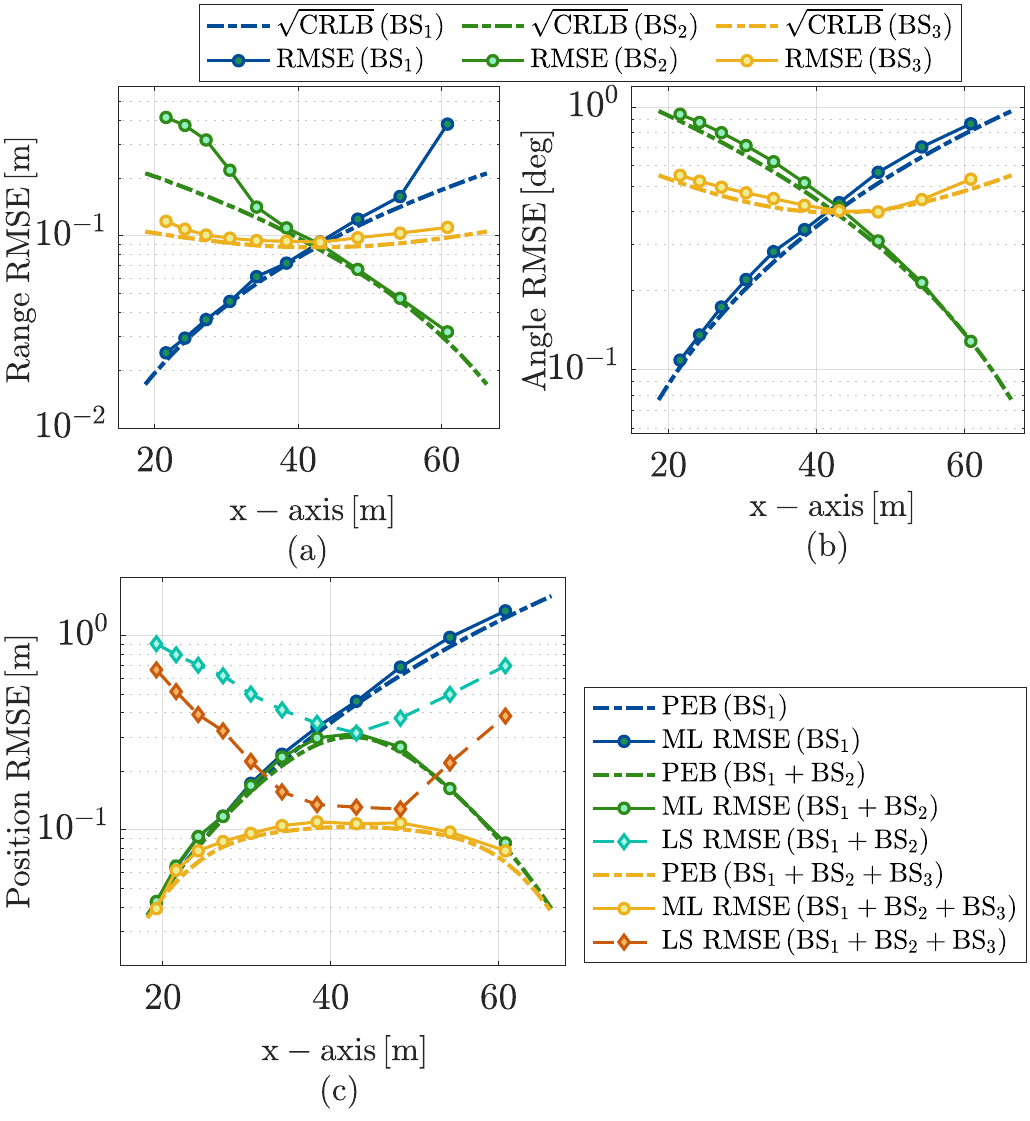}
     \caption{\ac{RMSE} and $\sqrt{\mathrm{CRLB}}$ for estimates of range $\hat{r}$ (a), angle $\hat{\phi}$ (b), and position $\hat{\mathbf{p}}$ (c) as the target moves along the diagonal line between $\mathrm{BS}_1$ and $\mathrm{BS}_2$ within the square area shown in Fig.~\ref{fig:ISAC_OTFS_system_model}. The $x$-axis represents the target’s position -- in the common reference system -- along the diagonal, where $y = x$.} 
     \label{fig:RMSE}
 \end{figure}
The proposed cooperative \ac{ML} estimation framework is validated through numerical simulation, varying the number of \acp{BS}, $\Nbs$, up to $3$, in the scenario shown in Fig.~\ref{fig:ISAC_OTFS_system_model} with system parameters listed in Table~\ref{tab:sim_param}. Additionally, cooperative \ac{ML} estimation performance is compared against (a) a single-\ac{BS} setup, where localization relies on local range and angle estimates, and (b) cooperative localization based on \ac{LS}. For \ac{LS}, following the linearization of the relationship between range and the target position, as described in \cite{Pelka2015}, a system of linear equations is formed by incorporating both \ac{AoA} and range estimates collected from each \ac{BS}. The \ac{LS} solution is then obtained according to \cite[Equation (20)]{ding2023LsRssAoaBased}, with range replacing RSS measurements and $\mathbf{W} = \mathbf{I}_{2N_\mathrm{bs}-1}$, where $2N_\mathrm{bs}-1$ is the number of equations.
The three \acp{BS} are numbered as in Fig.~\ref{fig:ISAC_OTFS_system_model} and placed at the corners of a square area with a side length of $85\,$m, with the origin of the common reference system in $\boldsymbol{\mathcal{O}}_1$. The following rotation angles, $[\vartheta_1, \vartheta_2, \vartheta_3] = [\pi/4, 5\pi/4, 3\pi/4]$, are considered.
The point-like target moves along the diagonal connecting $\mathrm{BS}_1$ and $\mathrm{BS}_2$.\footnote{This trajectory illustrates the benefits of cooperative localization, as the target experiences varying distances to $\mathrm{BS}_1$ while progressively benefiting from $\mathrm{BS}_2$ and $\mathrm{BS}_3$. The advantages demonstrated here apply to any trajectory within the monitored area.} As outlined in Section~\ref{sec:numerical_ML}, a two-stage estimation, consisting of a coarse and a refined estimation within a defined \ac{RoI}, is performed. For the coarse estimation, the search steps $\delta f_\mathrm{D}$, $\delta \tau$, $\delta \phi$ are computed with $c_\tau = 1$, $c_{f_\mathrm{D}}' = c_\phi' = 0.25$, with $\Delta \Phi \simeq 0.1108\,$rad. The matrix $\boldsymbol{\Psi}$ used to compute $\mathbf{G}$ in \eqref{eq:ML_estim_multi_2} and \eqref{eq:ML_pos_estim_2} is calculated based on the approximation method illustrated in \cite{BacPucPaoGio:J24}, specifically using $N_\mathrm{lobe}=5$ with reference to \cite[Equation (44)]{BacPucPaoGio:J24}.

Sensing performance is evaluated in terms of \ac{RMSE} of range, angle, and position estimates. For a given vector of parameters $\mathbf{q}$, this is computed as
$\mathrm{RMSE} = \sqrt{\frac{1}{N_{\mathrm{MC}}}\sum_{m=1}^{N_{\mathrm{MC}}} \|\hat{\mathbf{q}}_m-\mathbf{q}\|^2}$, being $N_\mathrm{MC} = 500$ the number of Monte Carlo iterations.
The results are shown in Fig.~\ref{fig:RMSE}, where \acp{RMSE} alongside the corresponding \acp{CRLB} are plotted against varying target's position. Since the target moves along the square's diagonal $y=x$, its position in the common reference system is represented here by its $x$ coordinate only. Fig.~\ref{fig:RMSE}a and Fig.~\ref{fig:RMSE}b show the \acp{RMSE} of range and angle estimates for single-BS scenarios. Additionally, the blue curve in Fig.~\ref{fig:RMSE}c provides the position estimation \ac{RMSE} when $\mathrm{BS}_1$ operates alone. These results highlight the limitations of single-BS localization, where position \ac{RMSE} can reach up to $1.34\,$m.
Conversely, Fig.~\ref{fig:RMSE}c demonstrates the benefits of cooperative \ac{ML} position estimation, which outperforms both the single-\ac{BS} scenario and the \ac{LS} solution, particularly in low signal-to-noise ratio regions for at least one \ac{BS}. For $\Nbs \geq 2$, position \ac{RMSE} is at most $0.3\,$m and $0.1\,$m with two and three \acp{BS}, respectively, compared to $0.9\,$m and $0.7\,$m with \ac{LS}. As a further example, when $x=54.2\,$m, the \ac{RMSE} is $0.98\,$m, $0.16\,$m, and $0.097\,$m for $1$, $2$, and $3$ \acp{BS}, respectively, while for the \ac{LS}, it is $0.50\,$m and $0.22\,$m with $2$ and $3$ \acp{BS}, respectively. Notably, for the considered single-target scenario, the proposed cooperative \ac{ML} framework achieves performance that closely matches the corresponding \ac{PEB}, underscoring the effectiveness of cooperative \ac{ML} estimation. 
\section{Conclusion} \label{sec:conclusion}
This letter proposed a cooperative \ac{ML} framework for target position estimation in \ac{MIMO} \ac{OTFS}-based \ac{ISAC} networks with multiple cooperating monostatic \acp{BS}. Results demonstrated the significant advantages of inter-\ac{BS} cooperation: specifically, the cooperative approach achieved a position estimation error below $10\,$cm with three \acp{BS}, a sevenfold improvement over single \ac{BS} configurations. Furthermore, the proposed framework closely approached the theoretical \ac{PEB}, underscoring its efficiency.

\appendix
\section{Maximum Likelihood Estimation}
Considering a given \ac{BS} and omitting its index 
$i$, after a few manipulations, \eqref{eq:log-likelihood} takes on the form 
\begin{align} \nonumber
    \label{eq:log_likelihood_1Bs}
    l(\y;\thetab,\mathbf{x}) \approx -\frac{1}{\sigma^2_\nu}\left(\beta^2 \bigl\| \mathbf{G} \mathbf{x} \bigr\|^2 + \bigl\| \y \bigr\|^2 - 2 \beta \mathfrak{Re} \{e^{j\varphi} \y^\herm \mathbf{G} \mathbf{x}\}\right).
\end{align}
Now, ignoring those constant positive terms that do not contribute to the estimation, the log-likelihood reduces to
\begin{equation}
\tilde{l}(\y;\thetab,\mathbf{x}) = 2 \beta \mathfrak{Re} \{e^{j\varphi} \y^\herm \mathbf{G} \mathbf{x}\} - \beta^2 \bigl\| \mathbf{G} \mathbf{x} \bigr\|^2.
\label{eq:log_likelihood_1Bs_2}
\end{equation}
From \eqref{eq:log_likelihood_1Bs_2} the vector of unknown parameters $\thetab$ can be estimated by maximizing the log-likelihood as
\begin{equation}
\label{eq:ML_estim}
\widehat{\thetab}_\mathrm{ML} = \underset{[\beta, \varphi, f_\mathrm{D}, \tau, \phi]}{\arg\max} \; \tilde{l}(\y;\thetab,\mathbf{x}).
\end{equation}
Since $\beta > 0$, the phase $\varphi$ of the complex channel coefficient $h$ that maximizes $\tilde{l}(\y;\thetab,\mathbf{x})$ is $\hat{\varphi} = -\angle{\y^\herm \mathbf{G} \mathbf{x}}$ and by replacing it in \eqref{eq:log_likelihood_1Bs_2}, this can be rewritten as
\begin{equation}
\label{eq:log_likelihood_1Bs_3}
\tilde{l}(\y;\thetab,\mathbf{x}) = 2 \beta |\y^\herm \mathbf{G} \mathbf{x}| - \beta^2 \bigl\| \mathbf{G} \mathbf{x} \bigr\|^2.
\end{equation}
The value of $\beta$ which maximizes the likelihood function can be easily obtained 
and used for 
$\widehat{h} = \widehat{\beta} e^{j\widehat{\phi}}$, as follows
\begin{equation}
\label{eq:h_estim}
\widehat\beta = \frac{|\y^\herm \mathbf{G} \mathbf{x}|}{\bigl\| \mathbf{G} \mathbf{x} \bigr\|^2}, \qquad
\widehat{h} = \frac{\mathbf{x}^\herm \mathbf{G}^\herm \y}{\bigl\| \mathbf{G} \mathbf{x} \bigr\|^2}.
\end{equation}
Lastly, by plugging $\widehat\beta$ into \eqref{eq:log_likelihood_1Bs_3}, solving \eqref{eq:ML_estim} reduces to
\begin{equation}
\label{eq:ML_estim_1Bs}
\left[\widehat{f}_\mathrm{D}, \widehat{\tau}, \widehat{\phi}\right] = \underset{[f_\mathrm{D}, \tau, \phi] \in {\mathbb{R}^3}}{\arg\max} \; \frac{|\y^\herm \mathbf{G} \mathbf{x}|^2}{\bigl\| \mathbf{G} \mathbf{x} \bigr\|^2}.
\end{equation}
\ifCLASSOPTIONcaptionsoff
  \newpage
\fi



\balance
\bibliographystyle{IEEEtran}
\bibliography{IEEEabrv,bibliography}
\end{document}